# Deep Learning Order Parameter for Polymer Phase Transition

Debjyoti Bhattacharya and Tarak K Patra[*]

Department of Chemical Engineering
Indian Institute of Technology Madras, Chennai TN 600036, India

**Abstract**

We report a deep learning (DL) framework viz. deep autoencoder that autonomously discovers an appropriate order parameter from molecular dynamics (MD) simulation data to characterize the coil to globule phase transition of a polymer. The deep autoencoder encodes the 3N dimensional MD trajectory of a polymer in a one-dimensional feature space and, subsequently, decodes the one-dimensional feature to its original 3N dimensional polymer trajectory. The feature space representation of a polymer provides a new order parameter that accurately describes the coil to globule phase transition as a function of temperature. This method is very generic and extensible to identify flexible order parameters to characterize wide range of phase transitions that take place in polymers and other soft materials. Moreover, this MD-DL approach is computational very efficient than a pure MD based characterization of phase transition, and has potential implications in accelerating phase prediction.

Developing effective order parameters for accurately classifying structures and phases is vital for advancing the current understanding of soft materials. Many local order parameters have been established to analyse molecular-scale structures of a material system, such as bond orientational order parameters that are commonly used to analyse crystal structures viz., bcc and fcc, as well as local orders of liquids and glasses.[1] The bond orientational order parameter use spherical harmonics to identify local order and often encounters large fluctuation depending on the definition of neighbour particles[2,3] which determines the efficiency of such order parameters. The neighbour particles are decided based on various approaches including hard-codded distances, choosing first minimum of a radial distribution function or using a Vornoi diagram. Clearly, there is no universal approach to determine the neighbour particles, and it always requires a-priori knowledge of the local order while formulating the order parameter. Such fixed functional form based local order parameters might fail to capture unknow phases, especially, amorphous phases. Moreover, there is a lack of universal approach to characterize various phase transition in materials. To address this limitation, here we develop a deep learning approach that autonomously produces order parameters form the molecular conformations of a system without any human intervention.

To develop and test a deep learning framework for extracting order parameters, we consider a canonical example of coil-globule transition (CGT) of polymers. The CGT has long been studied due to its theoretical importance and diverse applications.[4] Molecular simulations - Monte Carlo (MC) and molecular dynamics (MD) methods with either implicit or explicit solvent models and mean filed level theories have been extensively used to understand and predict the nature of the CGT and its connection to various molecular scale parameters.[5–10] However, accurate prediction of the crossover point is challenging and requires generating large amount of data using advanced sampling methods. Further, characterizing CGT using traditional approaches such as bond orientational order parameters is challenging. Here, we conduct standard molecular dynamics simulations of a bead-spring polymer model that undergo coil-globule transition in an implicit solvent condition during a cooling cycle, and use the simulation data to train a deep autoencoder. The deep autoencoder extracts an essential feature of the polymer MD trajectory in a one dimensional representation. This one dimensional description of a polymer chain is found to be appropriate order parameter that describes the phase transition as a function of temperature. This order parameter has no fixed functional form and it is completely flexible. The works shows how flexible order parameter can be extracted autonomously by analyzing MD trajectory using DL method, and open up new opportunity to predict and characterize phase transition more efficiently and with limited amount of data.

In this model system, two adjacent coarse-grained monomers of a polymer is connected by the Finitely Extensible Nonlinear Elastic (FENE) potential[11] of the form $E = -\frac{1}{2}KR_0^2\left[1-\left(\frac{r}{R_0}\right)^2\right]$, where, $k = 30\epsilon/\sigma^2$ and $R_0 = 1.5\sigma$. Any two monomers of the polymer chain is interacted via the Lennard-Jones (LJ) potential of the form $V(r_{ij}) = 4\epsilon_{ij}\left[\left(\frac{\sigma}{r_{ij}}\right)^{12} - \left(\frac{\sigma}{r_{ij}}\right)^6\right]$. The $\epsilon_{ij}$ is the interaction energy between the two monomers $i$ and $j$. The size of all the monomers are σ. The LJ interaction is truncated and shifted at a cut-off distance $r_c = 2.5\sigma$ to represent attractive interaction among the monomers. The molecular dynamics of the polymer chain is simulated for a range of temperate that incorporate coil to globule transition. The initial configuration of a polymer chain is placed in a cubic simulation box of fixed volume and the simulation box is periodic in all three dimension. The equation of motions of the monomers are integrated using the Velocity Verlet algorithm with a timestep of 0.001τ, where $\tau = \sigma\sqrt{m/\epsilon}$ is the unit of time. The temperature of the systems are controlled using the Langevin thermostat within the LAMMPS MD simulation environment.[12] The simulation is conducted for polymers of chain length N=30, 100 and 200. For each cases, the system is initially equilibrated at a high temperature and slowly cooled down to a low temperature by step wise quenching. We run the simulation for 50 temperature, and in each step, the temperature is reduced by 0.04. For any given temperature, the system is equilibrated for $10^8$ MD steps followed by a very short production cycle. We collect 500 configurations with a frequency of $10^3$ MD steps for each temperature. A total of 20000 frames are collected during the temperature annealing process for each chain length for developing the deep autoencoder. The deep autoencoder is essentially an artificial neural network which is designed in such a way that it compresses the correlations in an input dataset to a lower-dimensional representation and consequently decompresses them back to their original correlations. This compression and decompression of polymer data and their correlations are achieved by connecting neurons in a bottleneck structure as shown in Figure1. This autoencoder has three components - an encoder, a decoder and a feature space. The encoder compresses the 3N dimensional data of a polymer of chain length N to a one-dimensional feature space representation, which is also known as latent space. The decoder expands the one-dimensional representation into their actual 3N dimensional representation. Therefore, such bottleneck

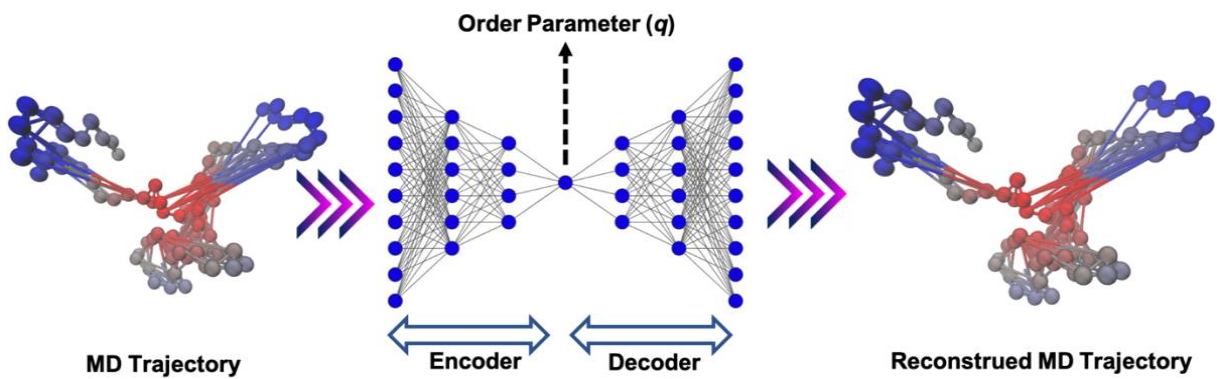

*Figure 1: Deep learning of polymer conformation. The autoencoder consists of two sub networks viz. encoders and decoder. The connections between a pair of neurons are shown by black lines. Each and every connection has a weight which is optimized during the training of the autoencoder. The encoder and decoder are connected by a single node that represents the characteristic order parameter of polymer conformations. The MD trajectory of a polymer is fed to the input layer of the autoencoder in the encoder section, the autoencoder reconstructs the MD trajectory at the output layer in the decoder section. er*

architecture allows automatic discovery of the features of a polymer configuration in low dimensional representation and helps classify the phases. Here, the topology of the network is $3N - n_1 - n_2 - n_3 - n_4 - 1 - n_4 - n_3 - n_2 - n_1 - 3N$, where each of these indices represents the number of neurons in the corresponding layers. Here, the number of neurons in the input and output layers are *3N* for a polymer of chain length *N*. The input layer and the output layer are identical as they hold the same position coordinates of a polymer chain. The $n_1, n_2, n_3$, and $n_4$ denote the numbers of neurons in the intermediate layers. Evidentially, the encoder and decoder part of the network are mirror images of each other. The middle layers has 1 neuron that represent the order parameter of the phase transition. Within this topology, each neuron of a given layer is connected to all the neurons of its adjacent layers by weight parameters. A neuron receives the weighted sum of signals from all the neurons of its previous layer and activate it by an activation function[13] and then feed it to all the neurons in its next layer. A rectified linear unit (ReLU) function[14] is used as a activation function for all the intermediate layer neurons. The output layer neuron activates the signal via a linear activation function and construct the polymer conformation. The network is trained by a feed-forward backpropagation method, wherein the weights between all the neurons' pairs are adjusted to minimize the error, which is a measure of the difference between the input values and the output values of the autoencoder. We use an Adam optimizer,[15] a first order gradient-based optimization method, with a learning rate of 0.001 to optimize weights of the network. A well trained autoencoder constructs a polymer conformation in its output layer which is identical to the one fed to the network via the input layer. More details of the training and development of a feed-forward backpropagation neural network can be found in our previous works.[16,17]

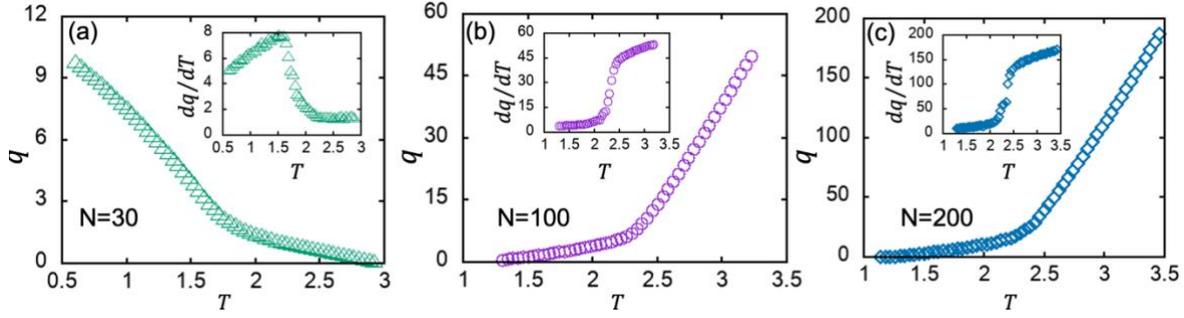

*Figure 2: The deep learning phase transition of polymer. The characteristic order parameter of a polymer chain (q) is plotted as a function of temperature for chain length N=30, 100 and 200 in (a), (b) and (c), respectively. The derivative of q with respect to T (dq/dT) is also shown in the inset of the respective figures.*

Once the autoencoder is trained, it is subjected to polymer conformations for estimating their projection to one dimensional feature space. The trained autoencoder produces an equivalent polymer trajectory at the output layer and its code values in the middle layer. This code vales in the middle layer representing the one dimensional mapping of the polymer configurations which is defined as the characteristic order parameter of a polymer chian. Therefore, the order parameter of the polymer chain during the phase transition can be written as $q = f\left(\sum_{m=1}^{n_4} a_5^m \cdot f_4^m \left(\sum_{l=1}^{n_3} a_4^l \cdot f_3^l \left(\sum_{k=1}^{n_2} a_3^k \cdot f_2^k \left(\sum_{j=1}^{n_1} a_2^j \cdot f_1^j \left(\sum_{i=1}^{3N} a_1^i \cdot x_i\right)\right)\right)\right)\right)$. Here, $x_i$s are the position coordinates of the polymer beads that are fed to the network. The weights between all the pairs of connected neurons are denoted by $a$, and $f(x) = max(0, x)$ is the rectified activation function (ReLU). The indices $i$, $j$, $k$, $l$, and $m$ account for number of neurons in 1st, 2nd, 3rd, 4th and 5th layer of the encoder, respectively. The Figure 2 shows the order parameter q as a function of temperature for three chain lengths N =30, 100 and 200. It clearly suggests an inflection of q as a function of temperature for all the chain lengths. We also calculate the first derivative of q with respect to temperature that are shown in the insets of Figure 2a, b and c for N=30, 100 and 200, respectively. It shows a discontinuity in the first derivative of q and we infer this deflection as the phase transition point. We estimate the critical temperature $T_c$=1.3, 2.1 and 2.4 for N=30, 100 and 200, respectively. This DL predicted Tc are in close agreement with pure molecular simulation with advanced sampling.[18]

There exist a set of fixed functional form based order parameter that are commonly used to study many liquid-solid and solid-solid phase transitions. These order parameters are not transferable to different class of phase transition and dynamical crossovers, and generalizing these order parameters is challenging. Moreover, defining such order parameter like quantity for transitions between amorphous phases such as coil-globule transition, glass transition and other complex transformation is challenging. In such cases, macroscopic properties like density and specific heat are computed for a wide range of temperature to identify phase transition and dynamical crossovers. This requires sampling millions of data points from the system's configuration space, which can be computationally very expensive. Here, we show that deep learning can address these limitations and provide flexible order parameter. The current MD-DL method takes 20000 data points to identify the crossover temperature. The current deep autoencoder autonomously discover order parameter for characterizing the phase transition. It mitigates the need for a priori construction of a suitable order parameter, simplify and streamlines the analysis of phase behavior. Therefore, the deep learning scheme will enable systematics, accurate and automatic mining of flexible order parameters for characterizing and predicting wide range of phase transformations and dynamical crossovers in soft matter.

**Conflicts of interest**

There are no conflicts of interest to declare.

**Acknowledgement**

The work is made possible by financial support from SERB, DST, Govt of India through a start-up research grant. TKP acknowledges ICSR, IIT Madras for the initiation, seed, and exploratory research grants. This research used resources of the Argonne Leadership Computing Facility, which is a DOE Office of Science User Facility supported under Contract DE-AC02-06CH11357. We also used the computational facility of the Center for Nanoscience Materials. Use of the Center for Nanoscale Materials, an Office of Science user facility, was supported by the U.S. Department of Energy, Office of Science, Office of Basic Energy Sciences, under Contract No. DE-AC02-06CH11357. We acknowledge the use of the computing resources at HPCE, IIT Madras.